\newlength\smallfigwidth
\newlength\figwidth
\documentclass[aps,twocolumn,floatfix,amsmath,amssymb,showpacs,showkeys,letterpaper]{revtex4}

\newcommand{\be}{\begin{equation}}
\newcommand{\ee}{\end{equation}}
\newcommand{\ba}{\begin{align}}
\newcommand{\ea}{\end{align}}
\newcommand{\bn}{\begin{eqnarray}}
\newcommand{\en}{\end{eqnarray}}

\newcommand{\ii}{{\rm i}}
\newcommand{\bsub}{\begin{subequations}}
\newcommand{\esub}{\end{subequations}}

\usepackage{graphicx}
\usepackage{amsmath}
\usepackage{hyperref}
\usepackage{bm}

\begin{document}


\title{Magnetic oscillation modes in square lattice artificial spin ice}
\author{Thomas D. Lasnier}
\email{tdlasnier@k-state.edu}
\affiliation{Department of Physics, Kansas State University, Manhattan, KS 66506-2601}
\author{G.\ M.\  Wysin}
\email{wysin@phys.ksu.edu}
\homepage{http://www.phys.ksu.edu/personal/wysin}
\affiliation{Department of Physics, Kansas State University, Manhattan, KS 66506-2601}

\date{January 9, 2020}
\begin{abstract}
{Small amplitude dipolar oscillations are considered in artificial spin 
ice on a square lattice in two dimensions.  
The net magnetic moment of each elongated magnetic island in the spin ice is assumed to 
have Heisenberg-like dynamics.
%
%
Each island's magnetic moment is assumed to be influenced by shape anisotropies and by the dipolar 
interactions with its nearest neighbors.
The magnetic dynamics is linearized around one of the ground states, leading to an $8\times 8$ 
matrix to be diagonalized for the magnetic spin wave modes.  
Analytic solutions are found and classified as antisymmetric and symmetric with regard to their
in-plane dynamic fluctuations.
Although only the leading dipolar interactions are included,  modes similar to these
may be observable experimentally.}
\end{abstract}
\maketitle

\section{Introduction: Square spin ice and its dynamics}
\label{intro}
Nanostructured arrays of thin elongated magnetic islands on a substrate, known as artificial spin ices,
have received a lot of theoretical and experimental interest because of their unique properties 
and possibilities for technological applications\cite{Ryzhkin05,Moessner06,Castelnovo08,Balents10}.
The magnetic islands possess an Ising-like dipole moment that tends to point in one of two
directions parallel to the long axis of the island. 
The arrays are manufactured in a desired geometry that has built-in frustration, where
all pairwise dipolar interactions cannot be simultaneously minimized \cite{Anderson56}.  
For square lattice artificial spin ice, the lowest energy configuration at a vertex
between four neighboring dipoles follows an \textit{ice rule}: two dipoles point inward
and two dipoles point outward at a vertex \cite{Wang06}. 
This leads to a doubly degenerate ground state as depicted in Fig.\ \ref{square-gs} where each 
vertex follows the ice rule, although it may be very difficult to achieve simply by cooling 
the sample \cite{Morgan11}.
Reversal of dipoles in a ground state leads to the generation of topological excitations that
resemble magnetic monopoles, and are connected by energetic string excitations 
\cite{Mol09,Mol10,Moller09,Morgan11}.
%

If only dipolar interactions are considered in Monte Carlo simulations for an Ising spin ice
model \cite{Mol09,Mol10,Silva12}, annealing of the system from high towards low temperature 
brings it to a ground state.
That approach leaves out the energy barriers involved in dynamic reversal.
Each island has a strong  easy-axis anisotropy that maintains the dipole's direction close 
to the island's long axis, as well as a strong easy-plane anisotropy maintaining
the dipole's direction near the plane of the substrate.
The energy associated with shape anisotropy of the islands is rather large compared to 
both the dipolar interactions and thermal energy scales \cite{Wysin+12,Wysin+13}.
This means that reversal of individual dipoles is difficult by thermal activation, because
some dipole reversals can be easily blocked by the anisotropy barriers,  making it 
difficult for the system to relax into a ground state \cite{Li10,Nisoli10} 
unless fields are applied.
%

The dynamics that is associated with lowest frequency spin waves is especially relevant 
for understanding the stability and signatures of different magnetic configurations
as well as transitions among configurations.
Here we consider the linearized dynamics out of a ground state configuration (sometimes
called a vortex state), where no monopole excitations are present.
Due to strong exchange interactions among the atomic spins within each island, we assume the
that the net island dipoles have nearly constant magnitude, while moving in an
anisotropy potential due to shape anisotropy, as considered in Ref.\ \cite{Wysin+12}.
Iacocca \textit{et al.}\ \cite{Iacocca+16} refer to this as a {\em macrospin} approximation, where
they used a semi-analytic approach including diagonalization and micromagnetics for finding various
modes of oscillation in artificial square spin ice.
Other studies of oscillation modes\cite{Gliga+13,Jung+16} have been carried out to demonstrate how 
the mode spectrum is affected by the presence or absence of topological excitations, such as monopoles.
Arroo \textit{et al.}\ \cite{Arroo+19} studied the connection between magnetic configuration and 
spin wave spectra using micromagnetics on a small number of islands.

The model used here for spin waves in artificial ice is simplified, however, it has the advantage 
of an entirely analytic solution, but it avoids the internal dynamics within individual islands.
Each dipole also interacts with its neighboring dipoles; for tractability only the nearest
neighbor dipolar interactions are included here.
%
The dipole moments behave with continuous dynamics, as Heisenberg-like magnetic moments that
can point in any direction, as considered in previous studies of thermally excited spin ice 
\cite{Wysin+13,Wysin+15}.
The small-amplitude spin wave deviations away from the ground state are considered,  
for at least two purposes: (1) as ground state signature, and (2) to indicate what applied 
field frequencies and wave vectors will reorganize a configuration.


The spin wave modes are determined as follows.
In Sec.\ \ref{model} the square lattice spin ice model is summarized. In Sec.\ \ref{dynamics} the 
dynamics for the nearest neighbor dipolar coupling is described.  
The system obtained is linearized in Sec.\ \ref{linear} and the details of the modes found are given.
Some excitation spectra for different model parameters are described in Sec.\ \ref{spectra},
and results are summarized and their importance is highlighted in Sec.\ \ref{conclude}.

\section{Artificial square lattice spin-ice model}
\label{model}
%
\begin{figure}
\includegraphics[width=1.2\smallfigwidth,angle=0]{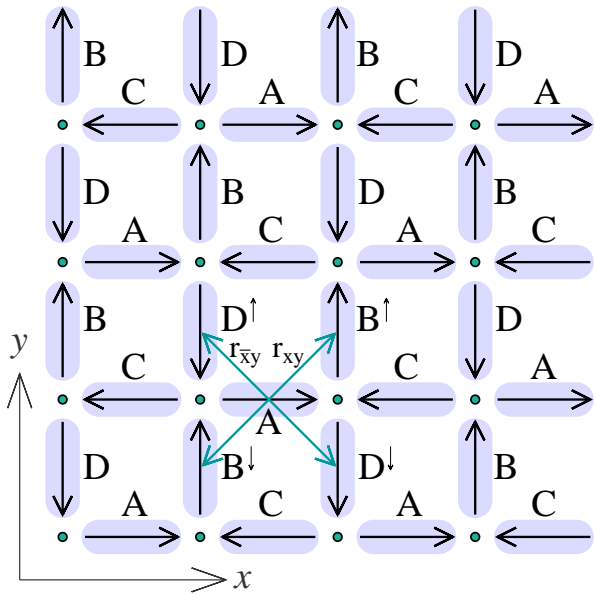}
\caption{\label{square-gs} Square spin ice in a ground state, with the identification of the four 
sublattices, for the four different directions of the islands' dipoles. Small dots indicate the 
vertices, at which the two-in/two-out rule holds. There are no monopole charges in this state. The
vertex lattice constant is $a$, while the nearest neighbor island spacing along diagonal directions 
is the island lattice constant $a_{\rm I}=a/\sqrt{2}$. For one A-site its nearest neighbors are labeled 
{\bf D}$^{\uparrow}$, {\bf D}$^{\downarrow}$, {\bf B}$^{\uparrow}$ and {\bf B}$^{\downarrow}$ and the 
displacements ${\bf r}_{\rm xy}$ and ${\bf r}_{\rm \bar{x}y}$ are indicated, see Eq. (\ref{Anbrs}).}
\end{figure}
The islands' dipoles are assumed to have fixed magnitudes $\mu$ pointing along some time dependent
Heisenberg-like unit vectors $\bm{\hat\mu}_i(t)$, where $i$ labels a site. The directions of the
$\bm{\hat\mu}_i(t)$ are affected by magnetic shape anisotropy and by long-range dipolar interactions.
Due to the elongated form of the islands, each island has some uniaxial anisotropy with energy constant
$K_1$ along its longer axis ${\bf \hat{u}}_i$, which points along either ${\bf \hat{x}}$ or ${\bf \hat{y}}$, 
depending on the sublattice.  A sketch of the system is shown in Fig.\ \ref{square-gs}.  In addition, 
the islands are very thin perpendicular to the substrate, which makes that direction a hard axis, 
producing easy-plane ($xy$) anisotropy with an energy constant $K_3$ for all the islands.  
In Ref.\ \cite{Wysin+12}, micromagnetics for an individual island indicates that the easy-plane 
anisotropy constant $K_3$ dominates, followed by the easy-axis interactions $K_1$, and then 
finally by the much weaker dipolar interactions.   Thermal energy scales can be expected
to be rather small compared to all of these couplings, which is why the system has a complex energy 
landscape with many local minima subject to frustration, typical of spin ice.   The Hamiltonian for 
this model with Heisenberg-like island spins $\bm{\hat\mu}_i(t)$ is
\bn
\label{Ham}
{\cal H} &=& -\frac{\mu_0}{4\pi} \frac{\mu^2}{a^3}
\sum_{i>j} \frac{ \left[ 3(\bm{\hat\mu}_i\cdot {\bf \hat{r}}_{ij})(\bm{\hat\mu}_j\cdot{\bf \hat{r}}_{ij})
                                -\bm{\hat\mu}_i\cdot \bm{\hat\mu}_j \right]}
{\left( {r}_{ij} / a\right)^3} \nonumber \\
&+&  \sum_{i} \left\{ K_1[1-(\bm{\hat\mu}_{i}\cdot{\bf \hat{u}}_i)^2] 
+ K_3 (\bm{\hat\mu}_{i}\cdot {\bf \hat{z}})^2 \right\}
\en
The first term is the dipolar pair interaction, where $\mu_0$ is the magnetic permeability of space, $a$ is
the center-to-center spacing of the islands along the ${\bf \hat{x}}$ or ${\bf \hat{y}}$ principal directions, 
and ${\bf \hat{r}}_{ij}$ is a unit vector pointing from site $j$ to site $i$.  Note, however, that the
nearest neighbor spacing of the islands, $a_{\rm I}=a/\sqrt{2}$, lies along the $\pm {\bf r}_{xy}$ and 
$\pm{\bf r}_{\bar{x}y}$ directions at $\pm 45^{\circ}$ from the standard $xy$ coordinate system, see Fig.\ 
\ref{square-gs}.  The dipolar energy scale is affected by island spacing, such that we define a nearest 
neighbor dipolar energy constant,
\be
\label{calD}
{\cal D} \equiv \frac{\mu_0}{4\pi} \frac{\mu^2}{a_{\rm I}^3}.
\ee
The anisotropy terms have been written so that they give zero
energy when the island dipole points along its local easy-axis ${\bf \hat{u}}_i$.  Rotation of $\bm{\hat\mu}_i(t)$
within the $xy$ plane only involves the $K_1$ energy, whereas, tilting of $\bm{\hat\mu}_i(t)$ out of the $xy$-plane
is characterized by the sum of the two anisotropy constants, $K_1+K_3$.

\subsection{The spin-ice ground states}
\label{ground}
In a ground state, such as in Fig.\ \ref{square-gs}, the shape anisotropy energies are totally 
minimized.  A ground state also does its 
best to minimize the nearest neighbor dipolar interactions, but those interactions are frustrated and
not globally minimized.  The magnetic moments alternate in direction from site to site, regardless of 
the displacement direction on the lattice.  
We use a notation where there are four sublattices, named A,B,C,D, as one moves
clockwise around a vertex where the ice-rule would be applied.  
The A and C sites are aligned with the $+{\bf \hat{x}}$ and $-{\bf \hat{x}}$ directions,
respectively, due to having in-plane anisotropy axes ${\bf \hat{u}}_i={\bf \hat{x}}$.
The B and D sites are aligned with the $+{\bf \hat{y}}$ and $-{\bf \hat{y}}$ directions,
respectively, due to having in-plane anisotropy axes ${\bf \hat{u}}_i={\bf \hat{y}}$.
In a ground state, the unit island dipoles $\bm{\hat{\mu}}_i$ on the different sublattices can be expressed as
\bsub
\label{gs}
\begin{align}
{\bf A}_0 & =  (\ 1,0,0), \quad \  {\bf B}_0 =  (0,1,0), \\
{\bf C}_0 & = (-1,0,0), \quad  {\bf D}_0 = (0,-1,0). 
\end{align}
\esub
This pattern repeats through the whole system, which then adheres to the ice rule throughout.
The other ground state would be obtained from this one by inverting all the moments.
There is an enormous energy barrier preventing that transition.
Instead, here we consider only small spatially periodic deviations away from this ground 
state configuration, characterized by some two-dimensional wave vector ${\bf q}=(q_x,q_y)$.

\section{The dynamics and symmetries}
\label{dynamics}
The dynamic equation of motion for the magnetic moment of some island, regardless of which
sublattice it occupies, results from the Hamiltonian according to a torque equation,
\be
\label{torq}
\frac{d\bm{\hat\mu}_i}{dt} = \gamma_{\rm e} \bm{\hat\mu}_i \times {\bf B}_i.
\ee
where $\gamma_{\rm e}$ is a gyromagnetic ratio.
Based on the local energies at each site, there is an effective magnetic field that acts on
the island at a site, 
\bn
\label{Bi}
{\bf B}_i &=& -\frac{\partial {\cal H}}{\partial \bm{\mu}_i} = -\frac{1}{\mu}\frac{\partial {\cal H}}{\partial \bm{\hat\mu}_i}
= \frac{\cal D}{\mu} \sum_{j\ne i} \frac{3 (\bm{\hat\mu}_j\cdot{\bf \hat{r}}_{ij}) {\bf \hat{r}}_{ij}-\bm{\hat\mu}_j}{(r_{ij}/a)^3}
\nonumber \\
&+& 2\frac{K_1}{\mu} (\bm{\hat\mu}_i\cdot {\bf \hat{u}}_i) {\bf \hat{u}}_i
- 2\frac{K_3}{\mu} (\bm{\hat\mu}_i\cdot {\bf \hat{z}}) {\bf \hat{z}} .
\en
In general, the anisotropy fields are local while the dipolar interactions extend through the entire lattice.

\subsection{Nearest neighbor dipolar model}
\label{nn-dipolar}
%
%
Although the dipolar interactions are long-ranged, in order to make initial progress and keep this calculation
tractable, only nearest neighbor dipolar couplings are included.  
The general properties of the solutions should not be significantly altered by this approximation.
%
To develop the equations for the undamped dynamics, we consider first a site on the A-sublattice,
and its interactions with the nearest neighbors on the B-sublattice and the D-sublattice, see Fig. \ref{square-gs}.
An arbitrary A-site couples to two B-sites whose unit dipoles are labeled as {\bf B}$^{\uparrow}$ and 
{\bf B}$^{\downarrow}$, and two D-sites whose unit dipoles are labeled as {\bf D}$^{\uparrow}$ and 
{\bf D}$^{\downarrow}$, where the arrows ($\uparrow, \downarrow$) indicate the y-direction of the space 
displacement from the A-site. 
To be specific, the displacements from the A-site to these neighbors are
\bsub
\label{Anbrs}
\begin{align}
{\bf r}_{\rm AB^{\uparrow}} & = {\bf r}_{\rm xy} \equiv (\ \ \tfrac{a}{2},\tfrac{a}{2},0), 
\quad {\bf r}_{\rm AB^{\downarrow}} = -{\bf r}_{\rm xy},  \\
{\bf r}_{\rm AD^{\uparrow}} & = {\bf r}_{\rm \bar{x}y} \equiv (-\tfrac{a}{2},\tfrac{a}{2},0), 
\quad {\bf r}_{\rm AD^{\downarrow}} = -{\bf r}_{\rm \bar{x}y}.  
\end{align}
\esub
These displacements have length $a'=a/\sqrt{2}$, which is the island lattice constant. 
From (\ref{torq}), the dynamic equation for the time derivative of the  A-site unit dipole 
$\bm{\hat\mu}_i \equiv {\bf A}$ can be expressed as
\be
\label{Adot}
\frac{d{\bf A}}{dt} = {\bf A}\times {\bf F}({\rm A}), 
\ee
where the effective field {\bf F}(A) acting on that site includes local anisotropy terms and only the nearest 
neighbor dipolar terms, 
\bn
\label{FA}
{\bf F}({\rm A}) & = & \kappa_1 A_x {\bf \hat{x}} -\kappa_3 A_z {\bf \hat{z}}  \\
 && + \delta_1   
\Big\{ 3\left[({\bf B}^{\uparrow}+{\bf B}^{\downarrow}) \cdot {\bf \hat{r}}_{\rm xy}\right] {\bf \hat{r}}_{\rm xy}
-{\bf B}^{\uparrow}-{\bf B}^{\downarrow}  \nonumber \\
 & & \quad \quad +  3\left[({\bf D}^{\uparrow}+{\bf D}^{\downarrow}) \cdot {\bf \hat{r}}_{\rm \bar{x}y}\right] {\bf \hat{r}}_{\rm \bar{x}y}
-{\bf D}^{\uparrow}-{\bf D}^{\downarrow}  \nonumber \Big\}.
\en
The constants $\kappa_1, \kappa_3$, and $\delta_1$ have dimensions of frequency and are defined as
\be
\label{defs1}
\kappa_1\equiv \frac{2\gamma_{\rm e} K_1}{\mu}, \quad \kappa_3\equiv \frac{2\gamma_{\rm e} K_3}{\mu}, \quad \delta_1\equiv\frac{\gamma_{\rm e} \cal D}{\mu}.
\ee
Once the nearest neighbor displacements are substituted into (\ref{FA}), the components of {\bf F}(A) are found to be
\bsub
\label{FAxyz}
\bn
F_x({\rm A}) &=&  \delta_1 \big[ \tfrac{1}{2}\left(B_x^{\uparrow}+B_x^{\downarrow}+D_x^{\uparrow}+D_x^{\downarrow}\right) \nonumber \\
& & + \tfrac{3}{2}\left(B_y^{\uparrow}+B_y^{\downarrow}-D_y^{\uparrow}-D_y^{\downarrow}\right) \big] +\kappa_1 A_x,\quad  \\
F_y({\rm A}) &=& \delta_1 \big[ \tfrac{1}{2}\left(B_y^{\uparrow}+B_y^{\downarrow}+D_y^{\uparrow}+D_y^{\downarrow}\right) \nonumber \\
& & + \tfrac{3}{2}\left(B_x^{\uparrow}+B_x^{\downarrow}-D_x^{\uparrow}-D_x^{\downarrow}\right) \big], \\
F_z({\rm A}) &=& -\delta_1\left(B_z^{\uparrow}+B_z^{\downarrow}+D_z^{\uparrow}+D_z^{\downarrow}\right) -\kappa_3 A_z.\ \
\en
\esub
By the symmetry of the lattice, a C-site follows a dynamic equation of the same form as in (\ref{Adot}) and (\ref{FA}), with 
the replacements ${\bf A}\rightarrow {\bf C}$, ${\bf B}\rightarrow {\bf D}$ and ${\bf D}\rightarrow {\bf B}$, and relations 
just like (\ref{Anbrs}) for the displacements:
\bsub
\label{Cnbrs}
\begin{align}
{\bf r}_{\rm CD^{\uparrow}} & = {\bf r}_{\rm xy} \equiv (\ \ \tfrac{a}{2},\tfrac{a}{2},0), 
\quad {\bf r}_{\rm CD^{\downarrow}} = -{\bf r}_{\rm xy},  \\
{\bf r}_{\rm CB^{\uparrow}} & = {\bf r}_{\rm \bar{x}y} \equiv (-\tfrac{a}{2},\tfrac{a}{2},0), 
\quad {\bf r}_{\rm CB^{\downarrow}} = -{\bf r}_{\rm \bar{x}y}.  
\end{align}
\esub
With these substitutions, a formula for effective field ${\bf F}({\rm C})$ is obtained from (\ref{FA}) and  (\ref{FAxyz})
with similar structure.

On the other hand, a B-site has two nearest neighbor A-sites with dipoles ${\bf A}^{\uparrow}$ and ${\bf A}^{\downarrow}$,
at displacements ${\bf r}_{\rm xy}$ and $-{\bf r}_{\rm xy}$, respectively, and two nearest neighbor C-sites with dipoles 
${\bf C}^{\uparrow}$ and ${\bf C}^{\downarrow}$, at displacements ${\bf r}_{\rm \bar{x}y}$ and $-{\bf r}_{\rm \bar{x}y}$, 
respectively.  With the B-site having a long axis along ${\bf \hat{y}}$, the effective field for its dynamics is
\bn
\label{FB}
{\bf F}({\rm B}) & = & \kappa_1 B_y {\bf \hat{y}} -\kappa_3 B_z {\bf \hat{z}}  \\
 && + \delta_1   
\Big\{ 3\left[({\bf A}^{\uparrow}+{\bf A}^{\downarrow}) \cdot {\bf \hat{r}}_{\rm xy}\right] {\bf \hat{r}}_{\rm xy}
-{\bf A}^{\uparrow}-{\bf A}^{\downarrow}  \nonumber \\
 & & \quad \quad +  3\left[({\bf C}^{\uparrow}+{\bf C}^{\downarrow}) \cdot {\bf \hat{r}}_{\rm \bar{x}y}\right] {\bf \hat{r}}_{\rm \bar{x}y}
-{\bf C}^{\uparrow}-{\bf C}^{\downarrow}  \nonumber \Big\}
\en
The Cartesian components now have the easy-axis anisotropy term in the y-component:
\bsub
\label{FBxyz}
\bn
F_x({\rm B}) &=&  \delta_1 \big[ \tfrac{1}{2}\left(A_x^{\uparrow}+A_x^{\downarrow}+C_x^{\uparrow}+C_x^{\downarrow}\right) \nonumber \\
& & + \tfrac{3}{2}\left(A_y^{\uparrow}+A_y^{\downarrow}-C_y^{\uparrow}-C_y^{\downarrow}\right) \big],  \\
F_y({\rm B}) &=& \delta_1 \big[ \tfrac{1}{2}\left(A_y^{\uparrow}+A_y^{\downarrow}+C_y^{\uparrow}+C_y^{\downarrow}\right) \nonumber \\
& & + \tfrac{3}{2}\left(A_x^{\uparrow}+A_x^{\downarrow}-C_x^{\uparrow}-C_x^{\downarrow}\right) \big] +\kappa_1 B_y,\quad \\
F_z({\rm B}) &=& -\delta_1\left(A_z^{\uparrow}+A_z^{\downarrow}+C_z^{\uparrow}+C_z^{\downarrow}\right) -\kappa_3 B_z.\ \
\en
\esub
Again by the symmetry of the lattice, the effective field ${\bf F}({\rm D})$ on a D-site is obtained from (\ref{FB}) or (\ref{FBxyz}) with 
the replacements ${\bf B}\rightarrow {\bf D}$, ${\bf A}\rightarrow {\bf C}$ and ${\bf C}\rightarrow {\bf A}$.  In this way, the general 
dynamics in the nearest neighbor dipolar approximation is fully described.

\section{Linearization around a ground state}
\label{linear}
Next we consider the small-amplitude magnetic fluctuations around the ground state defined in (\ref{gs}).  
To accomplish that, the four sublattices are assumed to have deviations from the ground state, denoted
as {\bf a, b, c, d}, with amplitudes much smaller than unity.   The net unit dipole fields are then
\bsub
\label{small}
\bn
{\bf A} &=& {\bf A}_0 + {\bf a} = (1+a_x,a_y,a_z), \\
{\bf B} &=& {\bf B}_0 + {\bf b} = (b_x,1+b_y,b_z),  \\
{\bf C} &=& {\bf C}_0 + {\bf c} = (-1+c_x,c_y,c_z), \\
{\bf D} &=& {\bf D}_0 + {\bf d} = (d_x,-1+d_y,d_z).  
\en
\esub
These can be used in the dynamic equations such as (\ref{Adot}) and its equivalent on the
other sublattices.
The equations are linearized, such that any terms quadratic and higher in these deviations are dropped.  
While the longitudinal deviations $a_x, b_y, c_x, d_y$ are included here, one finds after linearization 
that they all have zero time derivatives, $\dot{a}_x=\dot{b}_y=\dot{c}_x=\dot{d}_y=0$, so they can be assumed to 
be identically zero.    
Thus, the dynamic equations determine the time derivatives of the eight remaining fluctuation components,
that correspond to small-amplitude rotations of the islands' dipoles away from the ground state configuration.  
For example, on the A-sublattice one obtains from using (\ref{FAxyz}) in (\ref{Adot}) the results, 
\bsub
\label{adot}
\bn
\dot{a}_y & = & \delta_1 \left(6a_z+b_z^{\uparrow}+b_z^{\downarrow}+d_z^{\uparrow}+d_z^{\downarrow}\right) \nonumber \\
&&  +\kappa_{13} a_z,  \\
\dot{a}_z & = & \delta_1\left[-6a_y+\tfrac{3}{2}\left(b_x^{\uparrow}+b_x^{\downarrow}-d_x^{\uparrow}-d_x^{\downarrow}\right)\right] \nonumber \\
&&   -\kappa_1 a_y. 
\en
\esub
The combination of anisotropy constants appears,
\be
\kappa_{13}\equiv \kappa_1+\kappa_3. 
\ee
There are equations of similar structure for the other dynamically fluctuation pairs of components,
$(b_x,b_z)$, $(c_y,c_z)$, and $(d_x,d_z)$.  
On the C-sites, due to its ground state direction being reversed compared to the A-sites, there are
sign reversals on the dipolar terms:
\bsub
\label{cdot}
\bn
\dot{c}_y & = & -\delta_1 \left(6c_z+d_z^{\uparrow}+d_z^{\downarrow}+b_z^{\uparrow}+b_z^{\downarrow}\right) \nonumber \\
&&  +\kappa_{13} c_z,  \\
\dot{c}_z & = & -\delta_1\left[-6c_y+\tfrac{3}{2}\left(d_x^{\uparrow}+d_x^{\downarrow}-b_x^{\uparrow}-b_x^{\downarrow}\right)\right] \nonumber \\
&&   -\kappa_1 c_y. 
\en
\esub
The B-sites resemble A-sites but with opposite dipolar sign and different easy axis:
\bsub
\label{bdot}
\bn
\dot{b}_x & = & -\delta_1 \left(6b_z+a_z^{\uparrow}+a_z^{\downarrow}+c_z^{\uparrow}+c_z^{\downarrow}\right) \nonumber \\
&&  -\kappa_{13} b_z,  \\
\dot{b}_z & = & -\delta_1\left[-6b_x+\tfrac{3}{2}\left(a_y^{\uparrow}+a_y^{\downarrow}-c_y^{\uparrow}-c_y^{\downarrow}\right)\right] \nonumber \\
&&   +\kappa_1 b_x. 
\en
\esub
Finally, the D-sites have reversed ground state compared to B-sites, but similar local anisotropy terms:
\bsub
\label{ddot}
\bn
\dot{d}_x & = & \delta_1 \left(6d_z+c_z^{\uparrow}+c_z^{\downarrow}+a_z^{\uparrow}+a_z^{\downarrow}\right) \nonumber \\
&&  -\kappa_{13} d_z,  \\
\dot{d}_z & = & \delta_1\left[-6d_x+\tfrac{3}{2}\left(c_y^{\uparrow}+c_y^{\downarrow}-a_y^{\uparrow}-a_y^{\downarrow}\right)\right] \nonumber \\
&&   +\kappa_1 d_x. 
\en
\esub
%

\subsection{Traveling wave dynamic modes}
\label{modes}
The linearized equations can be solved by assuming traveling waves for the small-amplitude
fields. For example, on the B-sites, we take
\be
 b_x({\bf r},t) = b_x {\rm e}^{\ii({\bf q}\cdot {\bf r}-\omega t)},
\ee
where $b_x$ is a complex wave amplitude, ${\bf q}=(q_x,q_y)$ is a wave vector and $\omega$ is the frequency for that 
wave vector.  The equations contain combinations of the neighbors of a site, which have been labeled by up
($\uparrow$) and down ($\downarrow$) arrows.   As these are always along the displacements
${\bf r}_{\rm xy}$ and ${\bf r}_{\bar{x}y}$, one gets, for instance,
\bsub
\bn
b_x^{\uparrow}+b_x^{\downarrow} & = &  b_x {\rm e}^{\ii({\bf q}\cdot {\bf r}-\omega t)} \left( {\rm e}^{\ii{\bf q}\cdot {\bf r}_{\rm xy}} 
+ {\rm e}^{-\ii{\bf q}\cdot {\bf r}_{\rm xy}} \right), \\
d_x^{\uparrow}+d_x^{\downarrow} & = &  d_x {\rm e}^{\ii({\bf q}\cdot {\bf r}-\omega t)} \left( {\rm e}^{\ii{\bf q}\cdot {\bf r}_{\rm \bar{x}y}} 
+ {\rm e}^{-\ii{\bf q}\cdot {\bf r}_{\rm \bar{x}y}} \right).
\en
\esub
The phase factors are denoted as
\bsub
\label{phfacs}
\bn
u &\equiv & {\rm e}^{\ii{\bf q}\cdot {\bf r}_{\rm xy}} + {\rm e}^{-\ii{\bf q}\cdot {\bf r}_{\rm xy}} = 2\cos[\tfrac{a}{2}(q_x+q_y)], \\
v &\equiv & {\rm e}^{\ii{\bf q}\cdot {\bf r}_{\rm \bar{x}y}} + {\rm e}^{-\ii{\bf q}\cdot {\bf r}_{\rm \bar{x}y}} 
= 2\cos[\tfrac{a}{2}(q_x-q_y)].
\en
\esub
This allows for a more concise representation of the linearized dynamic equations, which now becomes
an $8\times 8$ eigenvalue problem,
\bsub
\label{8x8}
\bn
-\ii\omega a_y &=& \delta_1(6 a_z+u b_z+v d_z)+\kappa_{13} a_z, \\
-\ii\omega a_z &=& \delta_1(-6 a_y+\tfrac{3}{2}u b_x-\tfrac{3}{2}v d_x)-\kappa_1 a_y, \\
-\ii\omega b_x &=& \delta_1(-6 b_z-u a_z-v c_z)-\kappa_{13} b_z, \\
-\ii\omega b_z &=& \delta_1(6 b_x-\tfrac{3}{2}u a_y+\tfrac{3}{2}v c_y)+\kappa_1 b_x, \\
-\ii\omega c_y &=& \delta_1(-6 c_z-u d_z-v b_z)+\kappa_{13} c_z, \\
-\ii\omega c_z &=& \delta_1(6 c_y-\tfrac{3}{2}u d_x+\tfrac{3}{2}v b_x)-\kappa_1 c_y, \\
-\ii\omega d_x &=& \delta_1(6 d_z+u c_z+v a_z)-\kappa_{13} d_z, \\
-\ii\omega d_z &=& \delta_1(-6 d_x+\tfrac{3}{2}u c_y-\tfrac{3}{2}v a_y)+\kappa_1 d_x. 
\en
\esub
Before eliciting a solution for the general eigenmodes, a physical analysis 
of the situation points towards the symmetry of the lowest frequency fluctuations.

\subsection{Lowest energy fluctuations}
\label{lowE}
\begin{figure}
\includegraphics[width=\smallfigwidth,angle=0]{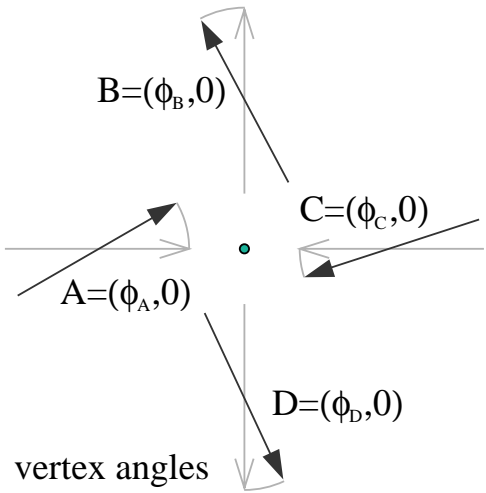}
\caption{\label{phiABCD} Deviations away from a ground state (faint gray arrows), for the in-plane dipolar angles 
in a vertex, drawn for a case where all of the angles $\phi_{\rm A}, \phi_{\rm B}, \phi_{\rm C}, \phi_{\rm D}$ 
are positive.  Out-of-plane components are ignored here.  These particular deviations tend to raise the 
in-plane nearest neighbor dipolar energies, see Eq.\ (\ref{Hvertex0}).} 
\end{figure}
We consider small angular fluctuations of the dipoles within the $xy$-plane, away from the ground state configuration.
A long wavelength mode is assumed to be present, wherein all the sites on a given lattice rotate nearly in-phase
with each other.
Consider the dipolar energy contributions around a single vertex of the lattice, see Fig.\ \ref{phiABCD}.  
Small in-plane angular deviations away from the ground state configuration are assumed, 
one for each sublattice: $\phi_{\rm A}, \phi_{\rm B}, \phi_{\rm C}, \phi_{\rm D}$.
Ignoring any small out-of-plane deviations, the unit dipole components for the sites on
the different sublattices in one vertex as in Fig.\ \ref{phiABCD} are
\bsub
\label{phi-ABCD}
\bn
{\bf A} &=& (\ \cos\phi_{\rm A}, \sin\phi_{\rm A},0), \\ 
{\bf B} &=& (-\sin\phi_{\rm B}, \cos\phi_{\rm B},0), \\ 
{\bf C} &=& (-\cos\phi_{\rm C}, -\sin\phi_{\rm C},0), \\ 
{\bf D} &=& (\ \sin\phi_{\rm D},-\cos\phi_{\rm D},0).
\en
\esub
The AB in-plane dipolar energy in (\ref{Ham}) for one vertex is found to be
\bn
\label{E-AB}
{\cal H}_{\rm AB}^{\rm dip} & = &-\frac{\cal D}{2} \big\{ 
3\cos(\phi_{\rm A}+\phi_{\rm B})+\sin(\phi_{\rm A}-\phi_{\rm B}) \big\} \nonumber \\
& \approx & -\frac{\cal D}{2} \big\{ 3+(\phi_{\rm A}-\phi_{\rm B})-\tfrac{3}{2}(\phi_{\rm A}+\phi_{\rm B})^2 
\big\}. \ 
\en
In the sine term, increasing $\phi_{\rm A}$ moves the A-dipole towards the direction of the vector ${\bf r}_{\rm AB}$,
which lowers the energy, while increasing $\phi_{\rm B}$ moves the B-dipole away from the direction of ${\bf r}_{\rm AB}$,
raising the energy.
The in-plane dipolar energy in the BC interaction follows the same rules (positive $\phi_{\rm B}$ moves the B-dipole
to be more aligned with ${\bf r}_{\rm BC}$, lowering the energy), 
\bn
\label{E-BC}
{\cal H}_{\rm BC}^{\rm dip} & = &-\frac{\cal D}{2} \big\{ 
3\cos(\phi_{\rm B}+\phi_{\rm C})+\sin(\phi_{\rm B}-\phi_{\rm C}) \big\} \nonumber \\
& \approx & -\frac{\cal D}{2} \big\{ 3+(\phi_{\rm B}-\phi_{\rm C})-\tfrac{3}{2}(\phi_{\rm B}+\phi_{\rm C})^2 
\big\}. \ 
\en
The CD in-plane dipolar energy in the vertex is lowered for positive $\phi_{\rm C}$,
\bn
\label{E-CD}
{\cal H}_{\rm CD}^{\rm dip} & = &-\frac{\cal D}{2} \big\{ 
3\cos(\phi_{\rm C}+\phi_{\rm D})+\sin(\phi_{\rm C}-\phi_{\rm D}) \big\} \nonumber \\
& \approx & -\frac{\cal D}{2} \big\{ 3+(\phi_{\rm C}-\phi_{\rm D})-\tfrac{3}{2}(\phi_{\rm C}+\phi_{\rm D})^2 
\big\}. \ 
\en
Finally, the in-plane dipolar energy in the DA interaction is lowered for positive $\phi_{\rm D}$,
\bn
\label{E-DA}
{\cal H}_{\rm DA}^{\rm dip} & = &-\frac{\cal D}{2} \big\{ 
3\cos(\phi_{\rm D}+\phi_{\rm A})+\sin(\phi_{\rm D}-\phi_{\rm A}) \big\} \nonumber \\
& \approx & -\frac{\cal D}{2} \big\{ 3+(\phi_{\rm D}-\phi_{\rm A})-\tfrac{3}{2}(\phi_{\rm D}+\phi_{\rm A})^2 
\big\}. \ 
\en
Summing over the four nearest neighbor dipolar energy terms between AB, BC, CD, DA,  
leads to an expression with only quadratic terms,
\bn
\label{Hvertex0}
{\cal H}_{\rm vertex}^{\rm dip} \approx \frac{\cal D}{2} & \big\{ &
-12+\tfrac{3}{2} \big[(\phi_{\rm A}+\phi_{\rm B})^2+(\phi_{\rm B}+\phi_{\rm C})^2 \nonumber \\
&&+(\phi_{\rm C}+\phi_{\rm D})^2+(\phi_{\rm D}+\phi_{\rm A})^2 \big] \big\}
\en
Then, if the dipoles rotate in such a way to minimize ${\cal H}_{\rm vertex}^{\rm dip}$, 
the motion must be constrained according to the phase relationships,
\be
\label{out-phase}
\phi_{\rm A} = -\phi_{\rm B}, \ 
\phi_{\rm B} = -\phi_{\rm C}, \
\phi_{\rm C} = -\phi_{\rm D}, \
\phi_{\rm D} = -\phi_{\rm A}. 
\ee
This means that in a low-energy (or low-frequency) mode, neighboring dipoles will tend
to move \textit{out-of-phase} with each other. 
These equations also then imply an \textit{in-phase} relationship across the two diagonals of the vertex:
\be
\label{in-phase}
\phi_{\rm A} = \phi_{\rm C}, \quad
\phi_{\rm B} = \phi_{\rm D}. 
\ee
Taken together, these conditions  would be met, for instance, when in-plane deviations $\phi_{\rm A}$ and 
$\phi_{\rm C}$ are both positive, while $\phi_{\rm B}$ and $\phi_{\rm D}$ are both negative (or \textit{vice-versa}).
%

If the in-plane dipolar interactions were the only interactions in the system, such fluctuations would
correspond to an acoustic mode in the system, whose frequency goes to zero for zero wave vector.
Of course, this system also has anisotropy terms and dipolar interactions of the out-of-plane components,
which will give this mode of fluctuation a nonzero frequency. 
This type of mode should have a minimum frequency for zero wave vector but it will not
be at zero frequency.
It is expected to become an acoustic mode in the
limit of zero easy-axis anisotropy (but that would no longer be a model for spin ice).
A mode that has this property will be referred to as an \textit{acoustic-like} mode.

By using (\ref{phi-ABCD}) or referring to Fig.\ \ref{phiABCD}, the angle constraints (\ref{in-phase}) imply
that for the Cartesian components as in (\ref{small}) or especially in (\ref{8x8}), we have for 
these lowest frequency modes, \textit{antisymmetry} across the center of the vertex,
\be
\label{ay-cy}
a_y = -c_y, \quad  b_x = - d_x.
\ee
We call this mode type \textit{antisymmetric} or type $A$, referring to the \textit{in-plane} dipolar deviations
across the center of a vertex.  
On the other hand, the other angular constraints (\ref{out-phase}) imply for the Cartesian components
of nearest neighbor dipoles,
\be
\label{ay-bx}
a_y = b_x, \quad c_y = d_x.
\ee
For these antisymmetric modes, we combine the constraints on the in-plane deviations with a phase 
relation (\ref{az-cz}) below for the out-of-plane components that results from consideration of the 
precessional dipolar motions.

The linearized energy in a vertex also includes dipolar energy in the out-of-plane components, and
the anisotropy energy that was initially not taken into account in (\ref{Hvertex0}).
When those terms are included, the total energy change away from the ground state is found to be
\bn
\label{Hvertex}
{\cal H}_{\rm vertex} & \approx & \frac{\cal D}{2} \big\{  
-12+\tfrac{3}{2} \big[(\phi_{\rm A}+\phi_{\rm B})^2+(\phi_{\rm B}+\phi_{\rm C})^2 \nonumber \\
&+&(\phi_{\rm C}+\phi_{\rm D})^2+(\phi_{\rm D}+\phi_{\rm A})^2 \big] \nonumber \\
&+&(a_z+c_z)(b_z+d_z) \big\} \nonumber \\
&+&K_1(a_y^2+b_y^2+c_y^2+d_y^2) \nonumber \\
&+&(K_1+K_3)(a_z^2+b_z^2+c_z^2+d_z^2).
\en
The anisotropy terms produce a nonzero frequency even at small wave vector. 
More interesting are the dipolar terms involving the $z$-components, $(a_z+c_z)(b_z+d_z)$.  
Those can be zeroed out, but not necessarily minimized, by assuming opposite phases across the 
vertex:
\be
\label{z-180}
a_z = - c_z, \quad b_z = - d_z. 
\ee
However, another possibility that could give even a negative energy contribution is if the 
$z$-components are in-phase across the vertex,
\be
\label{z-0}
a_z =  c_z, \quad b_z =  d_z, 
\ee
together with an opposite phase relation such as $a_z=-b_z$.  
The selection of one of these possibilities is decided next by analyzing the precessional
spin dynamics.

\subsection{Low energy precessional motion}
\label{precess}
\begin{figure}
\includegraphics[width=\smallfigwidth,angle=0]{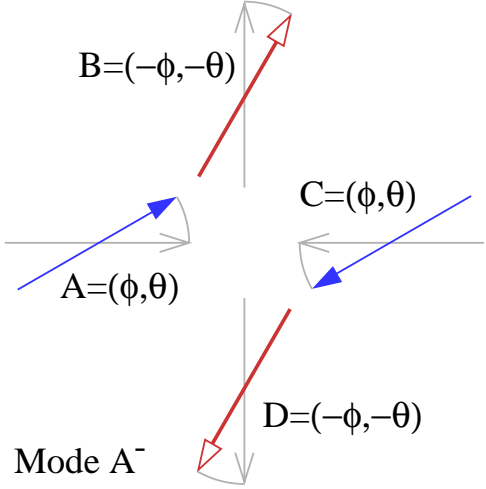}
\caption{\label{A-minus} Phase relationships of the dipolar angles expected in the antisymmetric
mode denoted as $A^{-}$, with in-plane rotations having $\phi_{\rm A}=-\phi_{\rm B}=\phi_{\rm C}=-\phi_{\rm D}$, 
and out-of-plane components obeying $a_z=-b_z=c_z=-d_z$.  A and C rotate in the same in-plane direction
and tilt positively out of plane together; B and D rotate together oppositely to A and C, and tilt out of 
plane together oppositely to A and C. These motions minimize the nearest neighbor dipolar energy changes,
see Eq.\ (\ref{Hvertex}).  This mode becomes acoustic-like in the limit of zero wave vector and zero
anisotropy.} 
\end{figure}

The choice of phase relationship for the $z$-components in a low energy mode was not determined
in the energy analysis.
Expressions (\ref{z-180}) and (\ref{z-0}) both appear to give low energy, without accounting
for the dynamics.
But some insight can be found by a comparison to the phase relationships that are present for 
spin wave modes in one-dimensional (1D) antiferromagnets, which require a two-sublattice model.  
Looking across a vertex, the A and C sites in the spin ice ground state alternate in direction
just as in a 1D antiferromagnet, which is known to have both acoustic and optical modes. 
The torque equation (\ref{torq}) shows that in a small time interval $\delta t$, the change in
the A-site dipole results from precession in the left hand sense around its effective field
${\bf F}({\rm A})$,
\be
\delta {\bf A} \approx {\bf A}\times {\bf F}({\rm A})\; \delta t.
\ee
From (\ref{FAxyz}), the effective field for the A-site is dominated by its $x$-component, 
\be
{\bf F}({\rm A}) \approx (6\delta_1 + \kappa_1, 0,  0).
\ee
With ${\bf A}\approx (1, a_y, a_z)$, this gives
\be
\delta {\bf A} \approx (6\delta_1 + \kappa_1)\delta t\, (0, a_z, -a_y).
\ee
By similar reasoning, a neighboring C-site precesses in the
left hand sense around its effective field, which is predominantly in the $-x$ direction, 
\be
{\bf F}({\rm C}) \approx (-6\delta_1 - \kappa_1,  0,  0).
\ee
With ${\bf C} \approx (-1,c_y, c_z)$, one has
\be
\delta {\bf C} \approx (6\delta_1 + \kappa_1)\delta t\, (0, -c_z, c_y).
\ee
From (\ref{ay-cy}) for low energy modes, using the relation $a_y=-c_y$ in the expression for $\delta{\bf C}$ gives 
\be
\delta {\bf C} \approx (6\delta_1 + \kappa_1)\delta t\, (0, -c_z, -a_y).
\ee
This shows that both the changes $\delta{\bf A}$ and $\delta{\bf C}$ across the center of a vertex
could have identical $z$-components for a low energy mode.  
Further, their $y$-components also are consistent with {\bf A} and {\bf C} having equal $z$-components.
Thus we should expect that any \textit{antisymmetric} mode should have an in-phase relation for 
the out-of-plane components:
\be
\label{az-cz}
a_z = c_z, \quad  b_z=d_z.
\ee
This should apply in conjunction with relations (\ref{ay-cy}) and (\ref{ay-bx}) for the in-plane 
components.
A sketch of the expected small deviations in one vertex for a lowest energy antisymmetric mode is given 
in Fig.\ \ref{A-minus}.
Both the A and C sublattices would rotate synchronized in-plane, in the same direction 
($\phi_{\rm A}=\phi_{\rm C}$), and they would also both tilt out of the $xy$-plane
with in-phase $z$-components.
The B and D sublattices would move together in the opposite sense compared to A and C, 
for both the in-plane and out-of-plane components.
These motions can be seen to minimize the linearized nearest neighbor dipolar energy changes within 
the vertex, see Eq.\ (\ref{Hvertex}).
These are the type of phase relationships present between the two sublattices in a 1D 
antiferromagnet for its lower frequency acoustic modes.
%


\subsection{Finding the antisymmetric modes}
For the antisymmetric modes, the fields on the C and D sublattices can be eliminated by imposing
the expected antisymmetric constraints from (\ref{ay-cy}) and (\ref{az-cz}), summarized together here:
\bsub
\label{A-assumption}
\bn
a_y =-c_y, &\quad& a_z=c_z, \\
b_x =-d_x, &\quad& b_z=d_z.
\en
\esub
Using this in the original $8\times 8$ system (\ref{8x8}) for only the A and B sublattices gives
\bsub
\label{ab-dot}
\bn
-{\rm i}\omega a_y &=& +(\kappa_{13}+6\delta_1)a_z+\delta_1 (u+v)b_z, \\
-{\rm i}\omega a_z &=& -(\kappa_1+6\delta_1)a_y+\tfrac{3}{2}\delta_1 (u+v)b_x, \\
-{\rm i}\omega b_x &=& -(\kappa_{13}+6\delta_1)b_z-\delta_1 (u+v)a_z, \\
-{\rm i}\omega b_z &=& +(\kappa_1+6\delta_1)b_x-\tfrac{3}{2}\delta_1 (u+v)a_y. 
\en
\esub
Subsequent equations will be simpler if new frequency constants are defined:
\bsub
\bn
\alpha_{1} &\equiv& \kappa_1+6\delta_1, \quad \alpha_{2}\equiv \kappa_{13}+6\delta_1,  \\
\gamma_{+} &\equiv& \delta_1(u+v) = 4 \delta_1 \cos(\tfrac{1}{2}q_x a) \cos(\tfrac{1}{2}q_y a).
\en
\esub
Applying another time derivative $d/dt = -{\rm i}\omega$ 
leads to two simplified systems where in-plane components are
separated from out-of-plane components.  
For the in-plane components, the dynamics obeys
\bsub
\label{aybx-dot}
\bn
\omega^2 a_y &=& +(\alpha_{1}\alpha_{2}+\tfrac{3}{2}\gamma_{+}^2)a_y-\gamma_{+}(\alpha_{1}+\tfrac{3}{2}\alpha_{2})b_x,  \\
\omega^2 b_x &=& -\gamma_{+}(\alpha_{1}+\tfrac{3}{2}\alpha_{2})a_y+(\alpha_{1}\alpha_{2}+\tfrac{3}{2}\gamma_{+}^2)b_x.
\en
\esub
For the out-of-plane components, the equations are nearly the same, except for a sign change on the
off-diagonal terms,
\bsub
\label{azbz-dot}
\bn
\omega^2 a_z &=& (\alpha_{1}\alpha_{2}+\tfrac{3}{2}\gamma_{+}^2)a_z+\gamma_{+}(\alpha_{1}+\tfrac{3}{2}\alpha_{2})b_z,  \\
\omega^2 b_z &=& \gamma_{+}(\alpha_{1}+\tfrac{3}{2}\alpha_{2})a_z+(\alpha_{1}\alpha_{2}+\tfrac{3}{2}\gamma_{+}^2)b_z.
\en
\esub
Both $2\times 2$ systems have the same eigenfrequencies, 
\be
\label{omA}
\omega_{A^{\pm}}^2 = (\alpha_{1}\alpha_{2}+\tfrac{3}{2}\gamma_{+}^2)\pm \gamma_{+}(\alpha_{1}+\tfrac{3}{2}\alpha_{2}).
\ee
The two frequencies $\omega_{A^{\pm}}$ correspond to two signs of the square root in the
eigenfrequency solution for these modes. 
A little consideration shows that for small wave vector $\omega_{A^-}$ is the lower of the two frequencies, 
and it goes to zero as $q\to 0$  when no uniaxial anisotropy is present ($\kappa_1=\kappa_3=0$).
The frequency $\omega_{A^+}$ tends to a large nonzero value at zero wave vector.
Note that Eq.\ (\ref{omA}) results in solutions for four of the eight possible modes of the original
$8\times 8$ system in Eq.\ (\ref{8x8}).
At a chosen wave vector {\bf q}, the possible frequencies are $\pm\omega_{\rm A^-}$ and $\pm\omega_{\rm A^+}$,
where the two signs relate to oppositely directed traveling waves that have the same absolute eigenfrequencies.

The modes' frequencies can also be written as the product of two factors:
\bsub
\label{omA+-}
\bn
\omega_{A^-}^2 & = & \left(\alpha_1-\tfrac{3}{2}\gamma_{+}\right)\left(\alpha_2-\gamma_{+}\right), \\
\omega_{A^+}^2 & = & \left(\alpha_1+\tfrac{3}{2}\gamma_{+}\right)\left(\alpha_2+\gamma_{+}\right).
\en
\esub
It is the factor $\left(\alpha_1-\tfrac{3}{2}\gamma_{+}\right)$ that tends to zero in the simultaneous
limit of zero wave vector and zero anisotropy, making it obvious that $\omega_{A^-}$ is an 
acoustic-like mode for this limit.

\subsubsection{Mode A$^-$ eigenvector and features}
For the mode at frequency $\omega_{A^-}$ we can also look at the structure of its eigenvector,
in terms of the phase relationships between the different dipolar components.
For its in-plane components, when the frequency $\omega_{A^-}$ is used in Eq.\ (\ref{aybx-dot}),
one immediately concludes that 
\be
a_y=b_x, \quad  c_y=d_x.
\ee
On the other hand, when the frequency $\omega_{A^-}$ is used in Eq.\ (\ref{azbz-dot}), it is
easy to see opposite phases for the out-of-plane components of neighboring dipoles, 
\be
a_z=-b_z, \quad c_z=-d_z.
\ee
This mode corresponds to angular deviations as represented in 
Fig.\ \ref{A-minus}. 
All of the in-plane angular deviations are of the same magnitudes, but with opposite phases 
between neighboring dipoles.
All of the out-of-plane deviations are also of equal magnitudes, but again with opposite phases
between neighboring dipoles.
For some eigenvector $\psi$, the deviations have pairs of in-plane and out-of-plane
Cartesian components on each sublattice, which we summarize in the following order:
\be
\label{psidef}
\psi = (a_y,a_z,\; b_x,b_z,\; c_y,c_z,\; d_x,d_z).
\ee
For this lowest antisymmetric mode (acoustic-like in the appropriate limit), the eigenvector of deviations 
in this notation is
\be
\psi_{A^-}=(a_y,a_z,\; a_y,-a_z,\; -a_y,a_z,\; -a_y,-a_z).
\ee
Therefore, the mode structure is determined by just two components.

The only other detail to consider, is how does $a_z$ compare in magnitude and phase to $a_y$?
That can be obtained by using $b_z=-a_z$ in Eq.\ (\ref{ab-dot}a), which results in
\be
a_z = \frac{-{\rm i}\omega_{A^-}}{(\alpha_2-\gamma_{+})} a_y
= -{\rm i} \left(\frac{\alpha_1-\tfrac{3}{2}\gamma_{+}}{\alpha_2-\gamma_{+}}\right)^{\frac{1}{2}} \; a_y.
\ee
One can see that in the acoustic-like limit of zero wave vector and zero anisotropy, $a_z$
tends towards zero, and the motion is predominantly in-plane.
%

\subsubsection{Mode A$^+$ eigenvector and features}
%
\begin{figure}
\includegraphics[width=\smallfigwidth,angle=0]{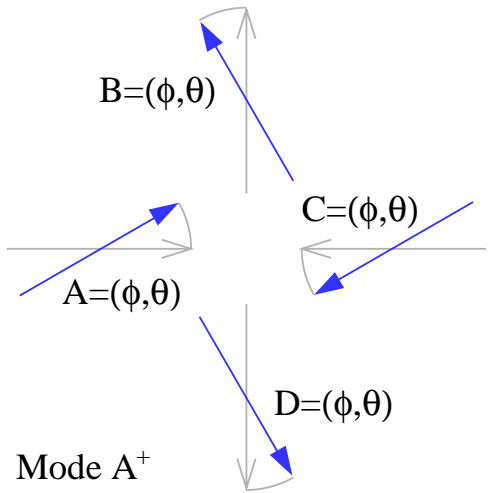}
\caption{\label{A-plus} Phase relationships of the dipolar angles expected in the antisymmetric mode 
denoted A$^+$, with frequency $\omega_{\rm A^+}$ given in Eq.\ (\ref{omA+-}b).
The in-plane rotations are equal and in-phase: $\phi_{\rm A}=\phi_{\rm B}=\phi_{\rm C}=\phi_{\rm D}$, and 
the out-of-plane components are also equal and in-phase: $a_z=b_z=c_z=d_z$.  
These motions tend to cause large changes in the nearest neighbor dipolar energies, see Eq.\ (\ref{Hvertex}).} 
\end{figure}
For the mode at the higher frequency, $\omega_{A^+}$, we expect different relative motions
of the sublattices.
For in-plane components, when frequency $\omega_{A^+}$ is used in Eq.\ (\ref{aybx-dot}),
we arrive at opposite phases for neighboring dipoles,
\be
a_y=-b_x, \quad  c_y=-d_x.
\ee
When combined with the assumptions in Eq.\ (\ref{A-assumption}), this shows that all
of the in-plane angles move together in-phase ($\phi_{\rm A}=\phi_{\rm B}=\phi_{\rm C}=\phi_{\rm D}$).
When the frequency $\omega_{A+}$ is used in Eq.\ (\ref{azbz-dot}), 
one also finds in-phase motions for the out-of-plane components,
\be
a_z=b_z, \quad c_z=d_z.
\ee
This implies then that all of the out-of-plane components move together in-phase, as well.
In the notation of Eq.\ (\ref{psidef}), the structure of Cartesian components for this mode is
\be
\psi_{A^+}=(a_y,a_z,\; -a_y,a_z,\; -a_y,a_z,\; a_y,a_z).
\ee
A sketch of this deviation structure is given in Fig.\ \ref{A-plus}.  
It is physically apparent that these angular deviations of the dipoles tend to raise
their nearest neighbor dipolar energy; this is not an acoustic-like mode in the limit
of zero anisotropy and wave vector.  
As far as the relative magnitudes of in-plane vs. out-of-plane components, we can use $b_z=a_z$
in Eq.\ (\ref{ab-dot}a) to arrive at the relation,
\be
a_z = \frac{-{\rm i}\omega_{A+}}{(\alpha_2+\gamma_{+})} a_y
= -{\rm i} \left(\frac{\alpha_1+\tfrac{3}{2}\gamma_{+}}{\alpha_2+\gamma_{+}}\right)^{\frac{1}{2}} \; a_y.
\ee
In the limit of zero wave vector and zero anisotropy, one finds that the $a_z$ and $a_y$ components
have similar magnitudes.
%

\subsection{Finding the symmetric modes}
Contrary to the assumptions made in Eq.\ (\ref{A-assumption}) for the antisymmetric modes, 
it is reasonable to assume that there are modes whose in-plane Cartesian components are 
\textit{symmetric} viewed across the center of a vertex, 
\bsub
\label{S-assumption}
\bn
a_y =c_y, &\quad& a_z=-c_z, \\
b_x =d_x, &\quad& b_z=-d_z.
\en
\esub
These are the same phase relationships that hold in the optic modes of a 1D antiferromagnet.  
They are assumed, however, it is straightforward to show that they do indeed
lead to solutions of the original $8\times 8$ system in Eq.\ (\ref{8x8}).

Using (\ref{S-assumption}) to eliminate the C and D sublattices, there results from (\ref{8x8})
the reduced $4 \times 4$ system,
\bsub
\label{Sab-dot}
\bn
-{\rm i}\omega a_y &=& +(\kappa_{13}+6\delta_1)a_z+\delta_1 (u-v)b_z, \\
-{\rm i}\omega a_z &=& -(\kappa_1+6\delta_1)a_y+\tfrac{3}{2}\delta_1 (u-v)b_x, \\
-{\rm i}\omega b_x &=& -(\kappa_{13}+6\delta_1)b_z-\delta_1 (u-v)a_z, \\
-{\rm i}\omega b_z &=& +(\kappa_1+6\delta_1)b_x-\tfrac{3}{2}\delta_1 (u-v)a_y.
\en
\esub
This suggest the definition of another wave vector dependent factor,
\be
\label{gamma-}
\gamma_{-} \equiv \delta_1 (u-v) = -4 \delta_1 \sin(\tfrac{1}{2}q_x a) \sin(\tfrac{1}{2}q_y a).
\ee
This factor becomes identically zero if $q_x=0$ or $q_y=0$.  Thus, the only symmetric modes
that will have some wave vector dependent features will not have wave vector aligned
with one of the lattice axes.

Taking the next time derivative of Eqs.\ (\ref{Sab-dot})  leads to separated
systems for the in-plane and out-of-plane components.  For in-plane, there results:
\bsub
\label{S-aybx-dot}
\bn
\omega^2 a_y &=& +(\alpha_{1}\alpha_{2}+\tfrac{3}{2}\gamma_{-}^2)a_y-\gamma_{-}(\alpha_{1}+\tfrac{3}{2}\alpha_{2})b_x,  \\
\omega^2 b_x &=& -\gamma_{-}(\alpha_{1}+\tfrac{3}{2}\alpha_{2})a_y+(\alpha_{1}\alpha_{2}+\tfrac{3}{2}\gamma_{-}^2)b_x.
\en
\esub
For out-of-plane, there is a sign change on the off-diagonal terms,
\bsub
\label{S-azbz-dot}
\bn
\omega^2 a_z &=& (\alpha_{1}\alpha_{2}+\tfrac{3}{2}\gamma_{-}^2)a_z+\gamma_{-}(\alpha_{1}+\tfrac{3}{2}\alpha_{2})b_z,  \\
\omega^2 b_z &=& \gamma_{-}(\alpha_{1}+\tfrac{3}{2}\alpha_{2})a_z+(\alpha_{1}\alpha_{2}+\tfrac{3}{2}\gamma_{-}^2)b_z.
\en
\esub
These are seen to be the same form as for the antisymmetric modes, but with the replacement $\gamma_{+}\rightarrow\gamma_{-}$.
Both $2\times 2$ systems have the same eigenvalues,
\bsub
\label{omS+-}
\bn
\omega_{S^-}^2 & = & \left(\alpha_1-\tfrac{3}{2}\gamma_{-}\right)\left(\alpha_2-\gamma_{-}\right), \\
\omega_{S^+}^2 & = & \left(\alpha_1+\tfrac{3}{2}\gamma_{-}\right)\left(\alpha_2+\gamma_{-}\right).
\en
\esub
This represents the four remaining modes of the original $8\times 8$ system. 
The factor $\gamma_{-}$ is nonzero only if both $q_x$ and $q_y$ are nonzero, 
and in the small wave vector limit, we have $\gamma_{-} \approx -q_x q_y a^2$.
These eigenfrequencies do not go to zero in the limit of zero wave length
and zero anisotropy.  
These modes have more of an optic-like character, with a finite frequency at zero wave vector
even in the limit of zero anisotropy.

\subsubsection{Mode S$^{-}$ eigenvector and features}
%
\begin{figure}
\includegraphics[width=\smallfigwidth,angle=0]{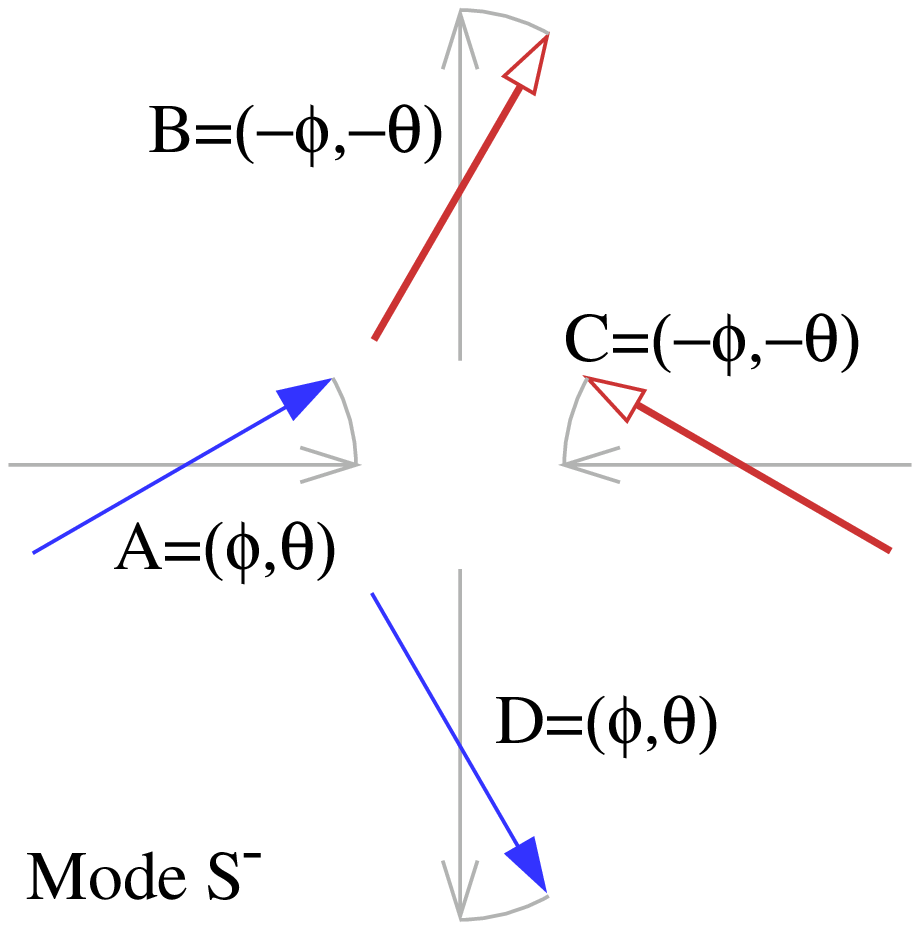}
\caption{\label{S-minus} Phase relationships of the dipolar angles expected in the symmetric mode 
denoted S$^-$, with frequency $\omega_{\rm S^-}$ given in Eq.\ (\ref{omS+-}a).
The in-plane angular deviations are towards the same side for dipole pairs across the vertex center.
The out-of-plane deviations are in opposite directions across the vertex center.  The
nearest neighbor relative deviations are partly energy reducing and partly energy enhancing.}
\end{figure}
For the mode with frequency $\omega_{S^-}$, substitution of the frequency into Eqs.\ (\ref{S-aybx-dot})
gives the relations,
\be
a_y = b_x, \quad c_y = d_x.
\ee
Using $\omega_{S^-}$ in Eqs.\ (\ref{S-azbz-dot}) leads to 
\be
a_z = -b_z, \quad c_z = -d_z.
\ee
These are the same nearest neighbor phase relations as for the mode A$^{-}$.  
Taken together with the symmetric assumption (\ref{S-assumption}), the eigenvector in
Cartesian components is of the form
\be
\psi_{S^-} = (a_y,a_z,\; a_y,-a_z,\; a_y,-a_z,\; a_y,a_z).
\ee
By using $a_z=-b_z$ in Eq.\ (\ref{Sab-dot}a), one arrives at the phase relation between in-plane
and out-of-plane components,
\be
a_z = \frac{-{\rm i}\omega_{S^-}}{(\alpha_2-\gamma_{-})} a_y
= -{\rm i} \left(\frac{\alpha_1-\tfrac{3}{2}\gamma_{-}}{\alpha_2-\gamma_{-}}\right)^{\frac{1}{2}} \; a_y.
\ee
A diagram of the deviations in a vertex is shown in Fig.\ \ref{S-minus}.  
Out of the four dipole-pair interactions, two of them reduce their energy while two of them increase 
their energy, compared to the ground state.
The AB and CD couplings move towards lower energy while the BC and DA couplings have moved towards 
higher energy.

\subsubsection{Mode S$^{+}$ eigenvector and features}
%
\begin{figure}
\includegraphics[width=\smallfigwidth,angle=0]{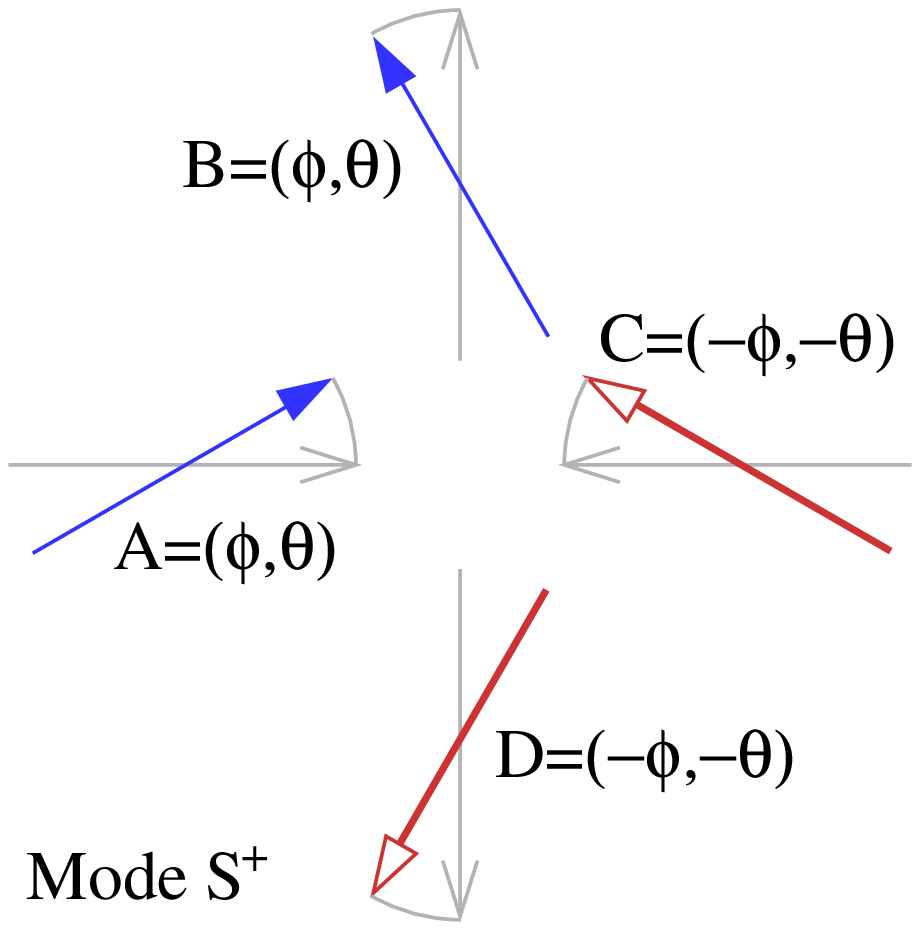}
\caption{\label{S-plus} Phase relationships of the dipolar angles expected in the symmetric mode
denoted S$^+$, with frequency $\omega_{\rm S^+}$ given in Eq.\ (\ref{omS+-}b).
The in-plane angular deviations are towards the same side for dipole pairs across the vertex center.
The out-of-plane deviations are in opposite directions across the vertex center.  The
nearest neighbor relative deviations are partly energy reducing and partly energy enhancing.}
\end{figure}
For the mode with frequency $\omega_{S^+}$, substitution of the frequency into Eqs.\ (\ref{S-aybx-dot})
gives the relations,
\be
a_y = -b_x, \quad c_y = -d_x.
\ee
Using $\omega_{S^+}$ in Eqs.\ (\ref{S-azbz-dot}) leads to
\be
a_z = b_z, \quad c_z = d_z.
\ee
These are the same nearest neighbor phase relations as for the mode A$^{+}$.
Together with the symmetric assumption (\ref{S-assumption}), the eigenvector in
Cartesian components is of the form
\be
\psi_{S^+} = (a_y,a_z,\; -a_y,a_z,\; a_y,-a_z,\; -a_y,-a_z).
\ee
By using $a_z=b_z$ in Eq.\ (\ref{Sab-dot}a), one arrives at the phase relation between in-plane
and out-of-plane components,
\be
a_z = \frac{-{\rm i}\omega_{S^+}}{(\alpha_2+\gamma_{-})} a_y
= -{\rm i} \left(\frac{\alpha_1+\tfrac{3}{2}\gamma_{-}}{\alpha_2+\gamma_{-}}\right)^{\frac{1}{2}} \; a_y.
\ee
A diagram of the deviations in a vertex is shown in Fig.\ \ref{S-plus}.  
In a certain sense it is very similar to the mode S$^-$.
Out of the four dipole-pair interactions, again two reduce their energy while two increase their energy.
The AB and CD couplings move towards higher energy while the BC and DA couplings have moved towards lower energy,
opposite to what takes place in mode S$^-$.
%

Indeed, there isn't a significant difference between modes S$^+$ and S$^-$, due to the behavior of
the factor $\gamma_{-}$, which reverses sign with a change in sign of either $q_x$ or $q_y$, see 
Eq.\ (\ref{gamma-}).
One can see $\omega_{S^-} \rightarrow \omega_{S^+}$ under a change such as $q_x\rightarrow -q_x$
or $q_y\rightarrow -q_y$.   
Thus, the two modes map into each other with an appropriate change of wave vector.

\section{Possible excitation spectra}
%
\label{spectra}
\begin{figure}
\includegraphics[width=\figwidth,angle=0]{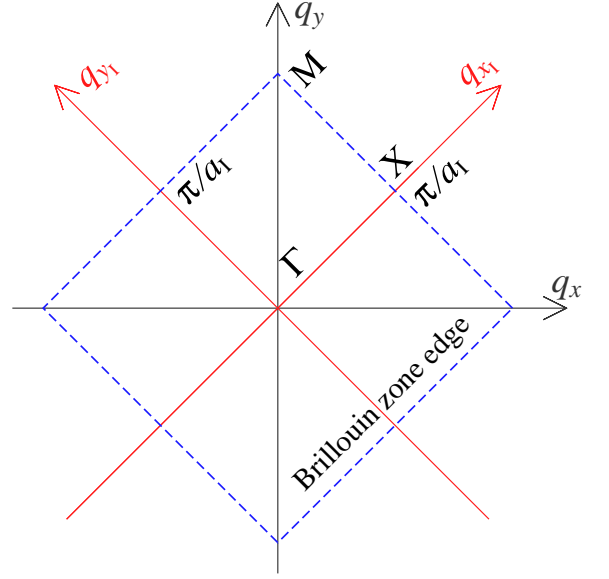}
\caption{\label{square-BZ} The first Brillouin zone for the square lattice of magnetic islands,
whose near neighbor spacing at 45$^{\circ}$ from the $x$-axis is $a_I=a/\sqrt{2}$.}
\end{figure}
Here we calculate some spectra for the excitations in a couple of situations.
The anisotropy constants $\kappa_1$, $\kappa_3$, and $\kappa_{13}$ as well as the dipolar coupling 
$\delta_1$ depend on the specific geometry of the islands.  
In a typical artificial spin ice, it is likely that the anisotropy constants dominate over the
dipolar coupling.  
Even so, it is instructive to consider some different choices of these parameters to observe how
they affect the mode frequencies.

For convenience here, frequencies will be measured in units of $\delta_1$.  We assume elliptical islands 
like those studied by Wang \textit{et al.} \cite{Wang06} with length $L_x=220$ nm, width $L_y=80$ nm and 
thickness $L_z=25$ nm.
If the material is Permalloy with saturation magnetization $M_s=860$ kA m$^{-1}$, the dipole moment
per island is $\mu=2.97\times 10^{-16}$ A m$^2$, see  Wysin \textit{et al.} \cite{Wysin+13}.  
We take a lattice constant $a=320$ nm, then the dipolar coupling constant from Eq.\ (\ref{calD}) is 
${\cal D} \approx 7.6\times 10^{-19}$ J.
Using the electron gyromagnetic ratio $\gamma_{\rm e}=1.76\times 10^{11}$ T$^{-1}$ s$^{-1}$, Eq.\ (\ref{defs1})
gives the value of the dipolar angular frequency constant, $\delta_1 \approx 4.5\times 10^{8}$ s$^{-1}$,
corresponding to a frequency unit $\delta_1/2\pi \approx 72$ MHz. 

\begin{figure}
\includegraphics[width=\figwidth,angle=0]{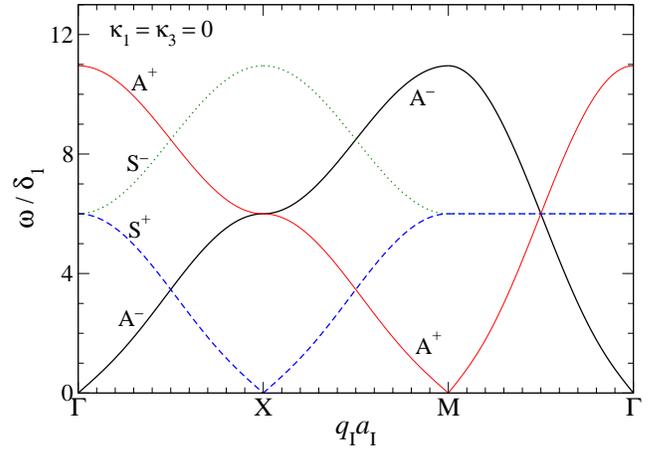}
\caption{\label{omega-kappa0} The excitation spectrum in the limit of zero anisotropy ($\kappa_1=\kappa_3=0$)
for wave vectors in the island coordinates from $\Gamma \to {\rm X} \to {\rm M}$ in the Brillouin zone,  
Fig.\ \ref{square-BZ}.  Modes S$^-$ and S$^+$ are degenerate along ${\rm M}\to\Gamma$. Mode A$^-$ is 
acoustic-like at ${\bf q}\to \Gamma$, while its sister-mode A$^+$ acquires zero frequency at the M-point.}
\end{figure}
%

The original $xy$ coordinate system was selected for finding the eigenmodes because the islands are 
oriented along those axes, however, the unit vectors of the island lattice are 
\be
\hat{x}_{\rm I} \equiv \tfrac{1}{\sqrt{2}}(\hat{x}+\hat{y}), \qquad 
\hat{y}_{\rm I} \equiv \tfrac{1}{\sqrt{2}}(-\hat{x}+\hat{y}). 
\ee
These are the directions of ${\bf r}_{xy}$ (45$^{\circ}$) and ${\bf r}_{\bar{x}y}$ (135$^{\circ}$) 
in Fig.\ \ref{square-gs}.
Then the dispersion relations for the modes should be calculated with wave vectors
${\bf q}=(q_{x_{\rm I}},q_{y_{\rm I}})$ expressed in this rotated coordinate system, within the first Brillouin zone,
as sketched in Fig.\ \ref{square-BZ}.  
Then the rotated components are
\be 
q_{x_{\rm I}} 
= \tfrac{1}{\sqrt{2}}(q_x+q_y), \qquad
q_{y_{\rm I}} 
= \tfrac{1}{\sqrt{2}}(-q_x+q_y). 
\ee
The phase factors used earlier in (\ref{phfacs}) are now simply $u=2\cos(q_{x_{\rm I}}a_{\rm I})$ and 
$v=2\cos(q_{y_{\rm I}}a_{\rm I})$, where $a_{\rm I}=a/\sqrt{2}$ is the near neighbor distance on the island lattice.  
This implies simplified phase factors in the dispersion relations,
\bsub
\bn
\gamma_{+} &=& \delta_1 (u+v)= 2\delta_1 [\cos q_{x_{\rm I}}a_{\rm I}+\cos q_{y_{\rm I}}a_{\rm I} ], \\
\gamma_{-} &=& \delta_1 (u-v)= 2\delta_1 [\cos q_{x_{\rm I}}a_{\rm I}-\cos q_{y_{\rm I}}a_{\rm I} ]. 
\en
\esub
These were used in dispersion relations (\ref{omA+-}) for A$^{\pm}$ modes and (\ref{omS+-}) for S$^{\pm}$ modes
to obtain the mode spectra for several situations.

\subsection{Zero anisotropy limit}
Initially, consider the extreme limit where the anisotropy constants are zero: $\kappa_1=\kappa_3=0$,
and only nearest neighbor dipolar coupling is present.
The resulting spectrum for the modes is shown in Fig.\ \ref{omega-kappa0},
with frequencies given in units of $\delta_1$.  
The antisymmetric mode A$^{-}$ is the acoustic-like mode, going to zero frequency linearly at zero wave vector.
The other antisymmetric mode, A$^{+}$, has its maximum frequency $\omega_{A^+}=\sqrt{120}\delta_1$ at ${\bf q}=0$ 
($\Gamma$), but acquires zero frequency at the M-point, where mode A$^{-}$ has its maximum frequency.
The symmetric modes are degenerate from M to $\Gamma$, or what corresponds to either having $q_x=0$ or $q_y=0$
in the original vertex coordinate system.
Along $\Gamma$ to X, however, the S$^+$ and S$^-$ frequencies move in opposite directions, with $\omega_{S^-}$
being higher.  
If one were to consider wave vectors from $\Gamma$ to Y (not shown), a similar structure would appear but with 
$\omega_{S^+}$ being higher.
As mentioned earlier, modes S$^{-}$ and S$^{+}$ map into each other, because the function $\gamma^{-}$ 
reverses sign if $q_x$ or $q_y$ is reversed in sign, which then takes $\omega_{S^+}$ into $\omega_{S^-}$
and \textit{vice-versa}.
Overall, one sees that there are several wave vector regions with a high density of low-energy modes present, 
of different symmetries.

\subsection{Weak island anisotropy}
%
\begin{figure}
\includegraphics[width=\figwidth,angle=0]{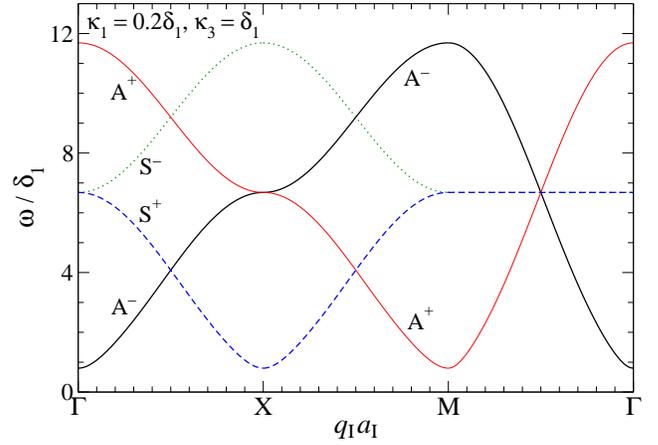}
\caption{\label{model015} The excitation spectrum for weak anisotropy, with $\kappa_1=0.2\delta_1$
and $\kappa_{3}=\delta_1$ for wave vectors in the first Brillouin zone of the island lattice.  
Note the small gap that opens up in the spectrum, of size
$\omega_{\rm gap}=\sqrt{\kappa_1(\kappa_{13}+2\delta_1)}=0.8\delta_1$.}
\end{figure}
Next, we suppose that the islands have weak shape anisotropies with energy
constants $K_1=0.1{\cal D}$ and $K_3=0.5{\cal D}$, but still with the same values of dipolar moment
$\mu=2.97\times 10^{-16}$ A m$^2$ and dipolar angular frequency $\delta_1=5.5\times 10^8$ s$^{-1}$
($f_1=\delta_1/2\pi=88$ MHz). 
Then the scaled anisotropy factors from Eq.\ (\ref{defs1}) are $\kappa_1=0.2\delta_1$ and
$\kappa_3=\delta_1$, which also gives $\kappa_{13}=1.2\delta_1$.
The mode spectrum that results is shown in Fig.\ \ref{model015}.
In the limit of small wave vector, a gap opens at ${\bf q}=0$ in the A$^{-}$ spectrum, given by
\be
\label{gap}
\omega_{\rm gap}=\omega_{\rm A^{-}}(0)=\sqrt{\kappa_1(\kappa_{13}+2\delta_1)}.
\ee
For the chosen parameters, the gap is $\omega_{\rm gap}=0.8\delta_1$.  
The same gap opens up for mode S$^{+}$ at X, mode S$^{-}$ at Y, and for mode A$^{-}$ at
the M points.
Now the acoustic-like mode is only weakly linear at long wavelength; the dispersion 
relations very near the frequency minima depend quadratically on the deviations of ${\bf q}$.

\subsection{Realistic anisotropy in a spin ice}
\begin{figure}
\includegraphics[width=\figwidth,angle=0]{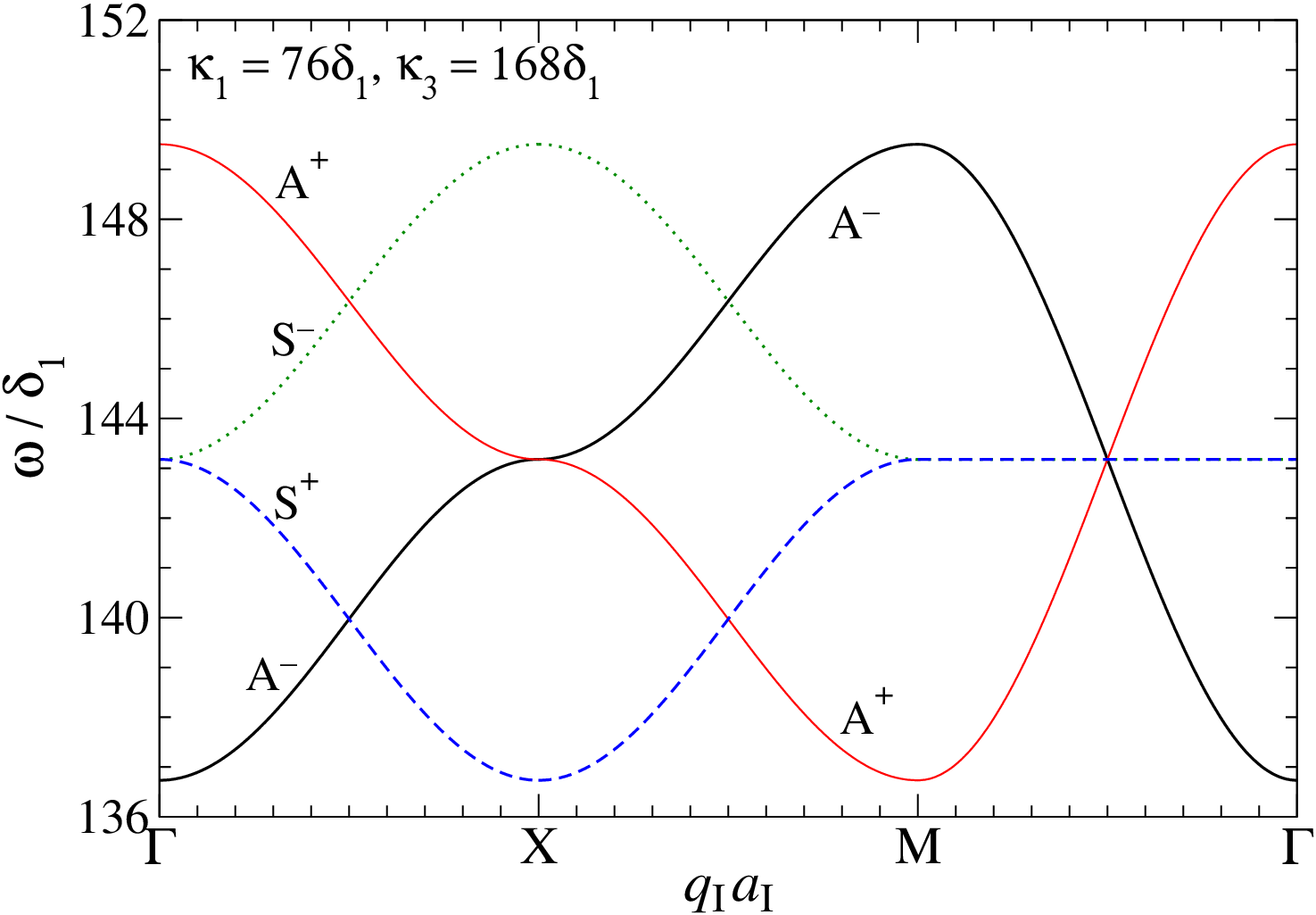}
\caption{\label{wang-GXM} The excitation spectrum for realistic anisotropy in a spin ice, 
with $\kappa_1=76\delta_1$ and $\kappa_{3}=168\delta_1$, for wave vectors in the first Brillouin
zone of the island lattice.  The spectrum is strongly elevated by a gap of size 
$\omega_{\rm gap}=\sqrt{\kappa_1(\kappa_{13}+2\delta_1)}=136.7\delta_1$,
but otherwise similar to that at weak anisotropy.}
\end{figure}

Finally it is important to show a prediction from this model for realistic parameters of
typical islands in artificial square spin ice, such as that studied by Wang \textit{et al.} \cite{Wang06}. 
Assuming elliptical islands with length $L_x=220$ nm, width $L_y=80$ nm and thickness $L_z=25$ nm,
energy minimization simulations indicate that their dipoles behave in a way described with easy-axis
anisotropy parameter $K_1\approx 2.9\times 10^{-17}$ J and hard-axis anisotropy parameter
$K_3=6.4\times 10^{-17}$ J.
For lattice parameter $a=320$ nm, we found above the dipolar energy constant ${\cal D} \approx 7.6\times 10^{-19}$ J. 
Then Eq.\ (\ref{defs1}) implies the anisotropy frequency constants are
\be
\kappa_1 \approx 76 \delta_1, \quad \kappa_3 \approx 168 \delta_1, \quad \kappa_{13}\approx 244 \delta_1.
\ee
As expected, the anisotropy is very strong compared to the dipolar interactions.   
This leads to a substantial gap in the spectrum,
\be
\omega_{\rm gap}=\sqrt{\kappa_1(\kappa_{13}+2\delta_1)} \approx 136.7\delta_1.
\ee
The resulting spectrum is shown in Fig.\ \ref{wang-GXM}.
One can see that the ${\bf q}$-dependence of the mode frequencies resembles that for weak anisotropy, except 
that the entire spectrum is elevated an amount equal to the gap frequency.
The variations in the mode frequencies with ${\bf q}$ are a rather small fraction of the total frequency. 
%

\section{Discussion and conclusions}
\label{conclude}
%
The eigenfrequencies and eigenvectors for four different types of modes have been found
analytically by diagonalization of the $8\times 8$ dynamic matrix for the model.
In the modes denoted as {\em antisymmetric}, the in-plane dipole components across the center of one 
vertex move oppositely.    
For mode A$^{-}$, both the in-plane and out-of-plane components of two nearest neighbor dipoles such as 
AB or AD also move oppositely relative to each other, see Fig.\ \ref{A-minus}.
To the contrary, for mode A$^{+}$, both the in-plane and out-of-plane components of two nearest neighbor 
dipoles move or rotate together in the same sense, see Fig.\ \ref{A-plus}.
For ${\bf q}\rightarrow 0$, the frequency of mode A$^{-}$ goes to a minimum; if
no anisotropy is present, that minimum frequency goes to zero linearly with ${\bf q}$, and 
mode A$^{-}$ is acoustic-like. 
An energy analysis for long wave vectors (Sec.\ \ref{lowE}) aided greatly in pointing towards 
the properties and phase relationships of the in-plane components of the mode that becomes 
acoustic-like.
An associated analysis of the precessional motion of a dipole (Sec.\ \ref{precess}) also was 
essential for understanding the phase relationships needed for the out-of-plane dipole
components for the lowest energy modes.
These symmetry considerations reduced the $8\times 8$ problem to smaller analytically tractable matrices.
%

In the other modes denoted as {\em symmetric}, the in-plane dipole components across the center of 
one vertex move in the same direction.
Depending on the choice of wave vector and especially its direction, one of the modes S$^{-}$
or S$^{+}$ may also go to low frequency in the limit of zero anisotropy.  
That is because their frequencies $\omega_{\rm S^{-}}$ and  $\omega_{\rm S^{+}}$ get interchanged
when the wave vector dependent factor $\gamma_{-}$ reverses sign, see Eq.\ (\ref{omS+-}).
This sign reversal would occur, for instance, by changing $q_x\rightarrow -q_x$ or
by $q_y\rightarrow -q_y$ (but not both together).
Indeed, a similar effect is present for the frequencies $\omega_{\rm A^{-}}$ and  $\omega_{\rm A^{+}}$
of modes A$^{-}$ and A$^{+}$, see Eq.\ (\ref{omA+-}), if the sign of the wave vector dependent
factor $\gamma_{+}$ is reversed.  
%

For nonzero anisotropy factors $K_1$ and $K_3$, a gap opens at the bottom of the spectrum, 
given by Eq.\ (\ref{gap});  mode A$^{-}$ acquires a finite frequency as ${\bf q}\rightarrow 0$.
The gap becomes significant for realistic anisotropy constants that might be expected for
typical elongated spin ice islands.
Still, there will be a {\bf q}-dependent modulation of the mode frequencies whose amplitude
depends on the nearest neighbor dipolar coupling, characterized by the dipolar frequency
$\delta_1$. 
%

There are two significant approximations used in this calculation: (1) that the island dipoles
essentially keep a constant magnitude $\mu$ but rotate uniformly, and (2) only nearest-neighbor 
dipolar interactions are included.
The first approximation is reasonable because only small-amplitude fluctuations are considered
for spin wave modes, and strong ferromagnetic exchange within the islands tends to preserve the
value of $\mu$.  
As a result, the spectra found here ignore magnetization dynamics within the islands, thus
the frequencies found here are higher than those in the semi-analytic calculations by
Iacocca \textit{et al.} \cite{Iacocca+16} and others \cite{Gliga+13,Arroo+19}.
We are not considering that any islands' dipoles rotate so far as to execute a reversal.
The nearest-neighbor approximation ignores the long range of dipolar interactions,
however, this facilitated the analytic solutions.
As a result, we cannot expect the dependence of frequency results on the dipolar
frequency $\delta_1$ (due to nearest neighbors only) to be completely correct.
The modes found should give some idea of the likely oscillatory motions, but 
the numerical details are approximate.
On the other hand, the dependencies of the mode frequencies on the anisotropy constants
such as $\kappa_1$ and $\kappa_3$, being local energy parameters, should be more reliable. 
Accounting for interactions beyond nearest neighbors will be the topic of a future study.

Ultimately, knowledge of the spin wave modes in artificial spin ice may be useful
for identifying differences between a ground and other states, for example.  
The presence of monopoles in excited states would modify the spectrum\cite{Gliga+13} as
the spin waves would scatter from monopoles.  
That effect is likely to broaden each mode frequency.
Calculations such as those presented here may be useful also for indicating the frequencies 
and polarization properties of applied magnetic fields intended to manipulate artificial 
spin ice states.


%
%
%


\end{document}